\documentclass[journal]{IEEEtran}
\ifCLASSINFOpdf
\else
\fi
\usepackage{cite}
\usepackage{graphicx}
\usepackage{epstopdf}
\usepackage{subfigure}
\usepackage[algoruled,linesnumbered]{algorithm2e}
\usepackage{threeparttable}
\usepackage{moreverb}
\usepackage{amsmath,amsthm, amssymb}
\usepackage{amsmath}
\usepackage{multirow}
\usepackage{chemarrow}
\usepackage{amssymb}
\usepackage{mhchem}
\usepackage{booktabs}
\usepackage{color}
\usepackage{makecell}
\usepackage{tabularx}
\begin{document}
\title{Novel  Authentication Protocols Tailored for Ambient IoT Devices in 3GPP 5G Networks}

\author{Xiongpeng~Ren,
        Jin~Cao,%~\IEEEmembership{Member,~IEEE,}
        Hui~Li,%~\IEEEmembership{Member,~IEEE,}
        and~Yinghui~Zhang%,~\IEEEmembership{Member,~IEEE,}% <-this % stops a space
\thanks{Xiongpeng~Ren, Jin Cao and Hui Li are with the State Key
	Laboratory of Integrated Services Networks, School of Cyber Engineering,
	Xidian University, Xi'an 710071, China. }% <-this % stops a space
\thanks{Jin Cao is the corresponding author. E-mail: caoj897@gmail.com.}

\thanks{Yinghui Zhang is with the National Engineering Laboratory for Wireless
	Security, Xi'an University of Posts and Telecommunications, Xi'an 710121,
	China, and also with Westone Information Industry INC., Westone Cryptologic
	Research Center, Beijing 100070, China.}}


\markboth{}%Journal of \LaTeX\ Class Files,~Vol.~14, No.~8, August~2015
{Shell \MakeLowercase{\textit{et al.}}: Bare Demo of IEEEtran.cls for IEEE Journals}



\maketitle

\begin{abstract}
AIoT devices have attracted significant attention within the 3GPP organization. These devices, distinguished from conventional IoT devices, do not rely on additional batteries or have extremely small battery capacities, offering features such as low cost, easy deployment, and maintenance-free operation. Authentication and secure transmission are fundamental security requirements for AIoT devices. However, existing standard security mechanisms are not specifically designed for AIoT devices due to their complex key hierarchies and multi-round interactions, making them unsuitable. Besides, AIoT devices would have more various communication topologies.
Therefore, we propose  dedicated ultra-lightweight access authentication protocols based on various technologies and algorithms  to serve as a forward-looking reference  for future research and standardization.  Analysis and simulation experiments using chips that closely resemble real AIoT devices, demonstrate that the existing standard protocols  are indeed not suitable for such devices, and
our protocols outperform existing standard protocols in terms of computational time and energy consumption.  After the successful  execution of proposed protocols, they can achieve secure transmission of application data, striking a balance between performance and security.

\end{abstract}


% \fi
\IEEEpeerreviewmaketitle

\section{Introduction}
\IEEEPARstart{A}{mbient}  power-enabled Internet of Things (IoT) devices, also known as passive IoT devices \cite{chinamobile}, or battery-free devices, are characterized by harvesting energy from the environment (such as Radio Frequency Energy, Solar Energy, Thermal Energy, etc.). These devices do not require additional  batteries or have very limited energy storage capability. They are highly cost-effective, easy to deploy, and require minimal maintenance. Additionally, they feature small form factors, low hardware complexity, and lower costs compared to existing IoT devices  (such as Redcap, NB-IoT, and eMTC) in 3GPP network.

In the past, significant research has been conducted on AIoT devices, with relevant companies conducting surveys on scenarios, energy harvesting technologies, and markets for  AIoT devices \cite{backscatter, market, oppowhitepaper}. In 2022, the 3GPP System Aspects1 (SA1) \cite{22840} group and Radio Access Network1 (RAN1) \cite{38848} group officially proposed many use cases and feasibility of  AIoT devices in Release 19. These terminals are suitable for four types of scenarios, including inventory management, sensor data collection, location tracking for easy tracing, and actuator control for signaling transmission. Given the devices' ultra-low cost and enormous potential market prospects, it is foreseeable that these devices have tremendous application potential. Therefore,  these devices have attracted significant attention.

The widespread deployment of  AIoT devices depends significantly on robust security measures. In particular, telecommunication networks must devise security access authentication and key agreement protocols specifically suited for AIoT devices. While existing security protocols \cite{33501, 33401, 33163} are in place, it's crucial to recognize that traditional authentication protocols were initially designed for smartphones and other more capable IoT terminals. Although these protocols provide strong security features, they may not be adequately suitable for AIoT devices due to  unique challenges of these devices, such as minimal or nonexistent batteries, extremely limited processing power, and constrained resources. Specifically, general devices must execute not only the primary authentication protocol with the network side, which leads to complex key  hierarchy, but also negotiate security abilities, security contexts at the Non-Access-stratum (NAS) and Access-stratum (AS) layers. These processes may significantly extend the operational time for AIoT devices to complete security protocols before  transmitting data. As a consequence, it results in higher power consumption. For instance, a device may fail to complete communication tasks before its energy is depleted, causing poor  availability. In addition, the AIoT devices foresee to have different kind of communication topologies.  Therefore, given that existing mechanisms are unsuitable,  new lightweight authentication protocols specifically for this type of device become necessary to strike a balance between security and efficiency.

 Our contributions in this paper can be summarized up as follows.

\begin{itemize}
\item We have proposed several lightweight authentication protocols utilizing various technologies such as squence numbers, nonces, and physical layer keys. The distinctive feature of AIoT devices, namely the absence of batteries, sets them apart from regular User Equipment (UE) and may lead to diverse communication topologies. Consequently, our authentication protocols are designed to accommodate different  topologies.
\item  The Ascon algorithm offers simultaneous encryption and authentication capabilities using one single key. It is a lightweight algorithm tailored for IoT devices and has been integrated into the NIST standard, showcasing vast potential for IoT applications. It is not excluded that in the future, 3GPP may adopt this new algorithm. Furthermore, it is preferable for authentication protocols to be compatible with the security algorithms in the existing 3GPP standards, avoiding significant modification costs to the current systems. Therefore, the protocol we propose supports various cryptographic algorithms.
\item Finally, we conducted informal security analysis for all proposed authentication protocols. The performance analysis of all authentication protocols was conducted. Through simulation experiments, it was found that the computational time of the current security protocols ranges from 2s to 6s. The computational time for  all of our protocols is under 1s. In terms of the energy overhead,  only proposed protocols  based on Ascon algorithm among all the schemes require the minimum power consumption to complete the entire authentication and secure transmission process. Besides, our protocols demonstrate a significant advantage compared to existing standard protocols, as indicated by the comparison results. Our work can  serve as a forward-looking reference  for future research and standardization implementation of AIoT devices.
\end{itemize}

The rest of the paper is organized as follows. A preliminary including  system architecture, Ascon algorithm, threat model  and security goal is presented in Section II. Section III describes the proposed protocols in detail. Security evaluation of the proposed scheme is discussed in Section IV including automated formal analysis and security property discussion. Section V presents the performance analysis among related protocols. Finally, the paper is concluded in Section VI.

\section{Related Work}

The 3GPP SA1 group launched a project in 2022 to investigate the use cases of AIoT devices and provided detailed service flows. This standard \cite{22840} will be active until December 2023. Concurrently, the RAN1 group \cite{38848} began exploring the feasibility of AIoT devices in 2022, classifying the use cases under \cite{22840} (e.g., indoor and outdoor, command, inventory, sensor, positioning) and outlining RAN design targets, along with a preliminary feasibility assessment. Furthermore, 3GPP proposals \cite{SA3115} focusing on AIoT device security  were presented in March 2024, hinting at the forthcoming security standards.

\begin{figure}[!t]
	\centering
	\includegraphics[width=3.5in]{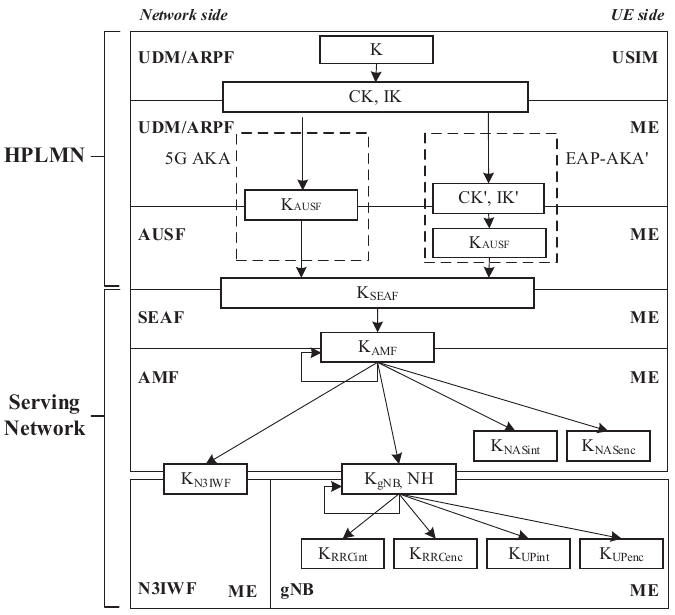}
	\caption{Key hierarchy of 5G AKA}
	\label{figure1}
\end{figure}

Currently, 3GPP standards encompass a variety of security authentication protocols tailored for different types of terminals \cite{33501, 33401, 33163}. These protocols include the 5G Authentication and Key Agreement (5G AKA) for general 5G UEs through 3GPP access, EAP-AKA' for non-3GPP accesses, EPS Authentication and Key Agreement (EPS AKA) for 4G terminals, the Control Plane Cellular Internet of Things (CP CIoT) security scheme for small data transmissions, and the Battery Efficient Security for very low Throughput Machine Type Communication (MTC) devices (BEST) authentication scheme specifically designed. The BEST authentication scheme can be implemented using various authentication protocols, including 5GAKA, EAP-AKA', EPSAKA, GBA, and AKMA. The authentication principles of these schemes are rooted in the  key K and the sequence number, employing the fi algorithm (it is essentially AES) and HMAC algorithm for implementation. These schemes feature different key derivation architectures. For instance, the key hierarchy of  5G AKA are depicted in Figure \ref{figure1}, illustrating its complexity. In these schemes, ordinary UEs are required to derive multiple types of keys, including NAS control plane keys, AS control plane keys, user plane keys, and intermediate keys. In the CP CIoT security scheme for small data transmissions, terminals can directly transmit small data over NAS CP, eliminating the need for generating access stratum control plane keys and user plane keys. The derived keys only consist of non-access stratum control plane keys and intermediate keys. In the BEST authentication scheme for MTC devices, MTC devices solely need to derive integrity and confidentiality keys between the device and the home network security endpoint, as well as intermediate keys for enterprise application services.

In the existing authentication schemes within the 3GPP standard, the computational overhead for security during the authentication process between terminals and the network side remains similar. This process involves performing 1 round of f1, f2 operations, and 1 round of HMAC operation (actually EAP-AKA' \cite{rfc5448} requires additional operations due to the presence of AT\_MAC). Different authentication schemes exhibit varying key derivation architectures. For instance, 5G AKA necessitates 1 round of f3, f4, f5 operations, and 10 rounds of HMAC-SHA256 operations, while EAP-AKA' requires 1 round of f3, f4, f5 operations, and 11 rounds of HMAC-SHA256 operations. Similarly, EPS AKA mandates 1 round of f3, f4, f5 operations, and 8 rounds of HMAC-SHA256 operations, while the CP CIoT security scheme demands 1 round of f3, f4, f5 operations, and 5 rounds of HMAC-SHA256 operations. Moreover, the BEST AKA, based on various protocols such as 5G AKA, EAP-AKA', EPS AKA, GBA, and AKMA, requires different numbers of HMAC-SHA256 operations ranging from 7 to 9 rounds. In cases where secure data transmission is necessary, the UE must transmit an extra signaling message and perform 1 round of encryption and integrity protection operations.

RFID can also be considered a type of AIoT devices that requires activation through a reader to obtain  information in the tag. This technology has been widely utilized and extensively researched in terms of security. For instance, in the \cite{RFIDonline}, the use of secret keys and the AES algorithm for authentication between the reader and RFID tags is mentioned. Additionally, some academic studies \cite{RFIDtdsc,fankai1,fankai2} have proposed alternative authentication methods. However, these methods often lack key negotiation functionality, and their security remains to be proven. Furthermore, these methods are not compatible with 3GPP network communication systems.

Due to the characteristics of AIoT devices, such as the absence of a battery or extremely small capacity battery, ultra-low power consumption, these terminals rely on harvesting energy from the environment to ensure their computing and normal communication operations. However, the current key derivation architecture of standard authentication protocols is relatively complex, and the computational power of terminals is extremely limited. Some keys (like $K_{SEAF}$, $K_{gNB}$) are meaningless. To achieve a balance between efficiency and security, it even may be possible to avoid using dual keys and instead use a single key for both encryption and integrity protection (According to the key usage guidance \cite{NISTkey}, two  encryption key and integrity key are seperated for security reasons, however,  keys are used to protect very small  data in the  scenarios for AIoT devices, meanwhile keys can be updated after each authentication). Based on the above statistical results, these solutions are likely unsuitable for resource-constrained AIoT devices. Therefore, it is necessary to propose an ultra-lightweight secure access authentication protocol suitable for such devices.
\begin{figure}[!t]
	\centering
	\includegraphics[width=3.5in]{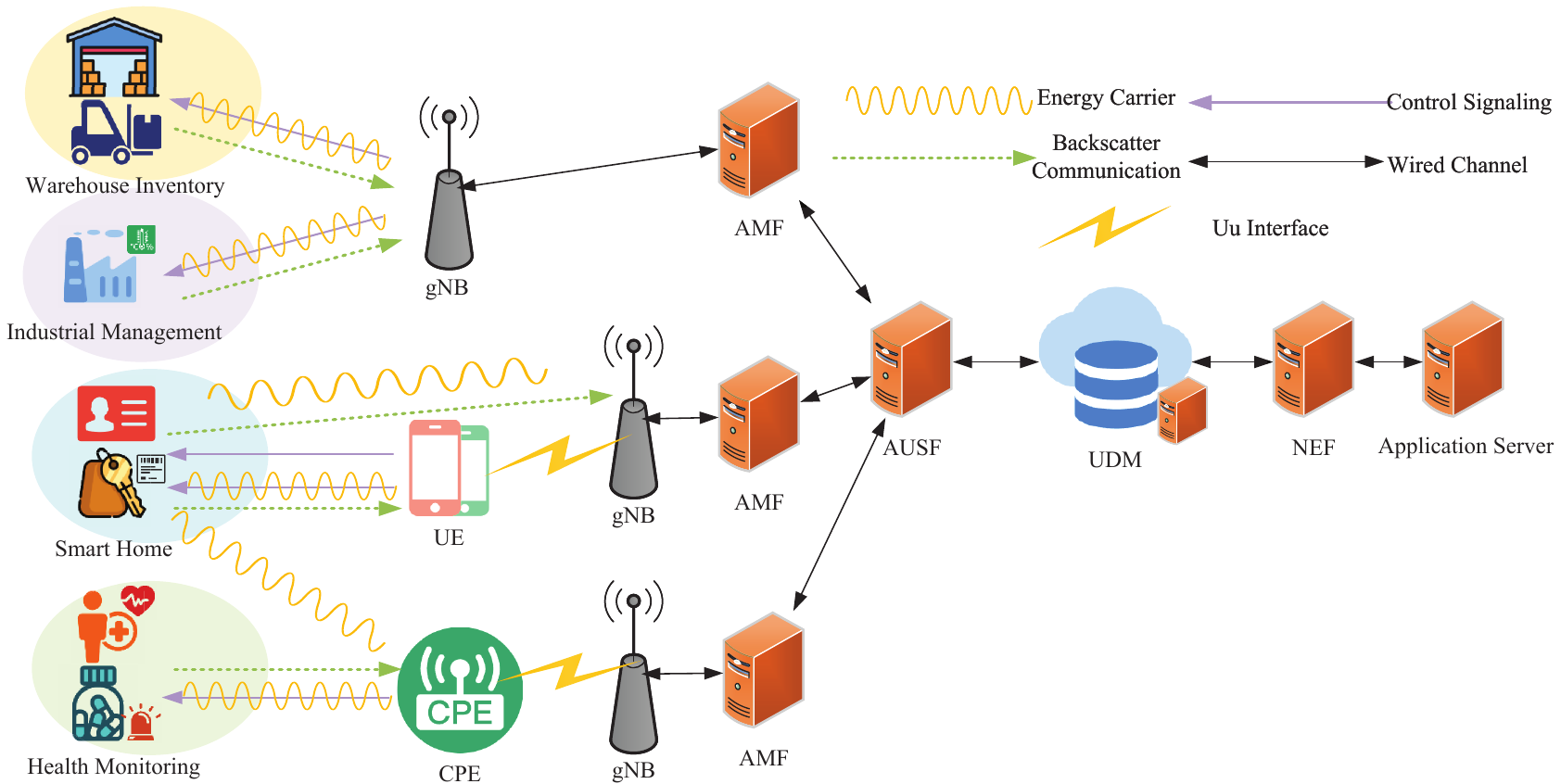}
	\caption{AIoT network architecture}
	\label{figure2}
\end{figure}

\section{Preliminary}

%\IEEEpubidadjcol
\subsection{AIoT network architecture}

As shown in Fig.\ref{figure2},  AIoT network architecture consists of four parts, including a variety of user devices (such as AIoT devices, UEs, Customer Premise Equipments (CPE)), gNBs, core network side (Access and Mobility Management Function (AMF), Authentication Server Function/Unified Data Management (AUSF/UDM), Network Exposure Function (NEF)), data network side. 
\begin{itemize}
	\item AIoT devices can be embedded in  the product itself and they can harvest energy from radio waves transmitted by gNBs or relays (such as smartphones, CPE). They  facilitate data transmission via backscatter signals. Some AIoT devices may also be equipped with supplementary  solid-state printed batteries \cite{22840}, and  these devices feature independent signal generation and amplification capabilities \cite{38848}.
	\item In the serving network, the AMF is responsible for the registration, connection, mobility management, and access authentication. 
	While in the home network, the UDM manages the storage of  identity, keys, nonces, and other subscription information shared with AIoT devices, the AUSF is responsible for access authentication within the home network. 
	\item  In the data network, the third appilication function can contact with the core network through NEF. 
\end{itemize}

The  standard TR 38.848 \cite{38848} mentions four types of connectivity topologies.  Firstly, the Ambient IoT device directly and bidirectionally communicates with a basestation. 
 The second topology involves AIoT devices engaging in bidirectional communication with base stations through intermediate nodes. The third topology is characterized by base stations directly transmitting downlink data to AIoT devices, which then relay uplink data to base stations via assisting nodes. Alternatively, base stations can transmit downlink data to AIoT through assisting nodes, after which AIoT devices directly send uplink data to base stations.  The intermediate nodes and assisting nodes both can be seen as a relay. The final topology is that UEs and AIoT devices communicate directly without network-side involvement. 

Based on the above connectivity topologies, we will integrate them into the following scenarios for consideration.
\begin{itemize}
	\item 
	Scenario 1 (Warehouse inventory): AIoT devices are deployed in a fixed area, allowing the network to locate the terminal. For instance, in a factory, when goods are being received or shipped out, base stations at the entrances need to scan a large number of devices to obtain information about the incoming goods. In this scenario, base stations deployed within the factory supply power to the devices and wake them up, enabling direct communication between the two.
	\item
	Scenario 2 (Industrial  sensing and managment): An important difference from Scenario 1 is that the network proactively sends  requests to specific devices. For instance, it may issue control commands to activate AIoT devices, or request sensor data from devices located in factories. In such instances, the network actively transmits authentication messages to the designated AIoT terminals, which respond to the network side. In this scenario, the network takes the lead by initiating the process of charging and waking up AIoT devices via base stations.
	\item 
	Scenario 3 (Smart home or health care): Unlike Scenarios 1 and 2, considering the limited coverage range of base stations (a coverage blind area exists), there are situations where the efficiency of energy supply is reduced due to the distance between terminals and base stations, resulting in lower reliability of the system. For instance,  in smart home scenarios where monitoring temperature/humidity or controlling devices intelligently is required,  or in health monitoring scenarios where medications are equipped with AIoT devices to remind patients to take specific drugs. In these scenarios, AIoT devices are within a certain range of user terminals, which may introduce user terminals (such as smartphones or CPEs) to supply energy and relay data. In this scenario, terminals transmit data upstream through relay nodes, while the network side transmits data downstream through relay nodes or directly.
	
	\item
	Scenario 4 (Smart home or health care): Unlike Scenario 3, irregular mobility of certain AIoT devices makes it challenging for the core network to determine which base station in the calling area should be used to wake up the terminals. Considering that such AIoT devices are often within a certain distance range of user terminals (such as smartphones or CPEs), the network side utilizes user terminals for positioning and waking up AIoT devices. In this scenario, the network transmits data downstream through relay nodes, while terminals transmit data upstream directly or through relay nodes.
\end{itemize}

\subsection{Threat Model}
We assume that adversaries can eavesdrop, intercept, replay, and tamper with messages in the wireless channel in an attempt to attack AIoT or the network. The connections between base stations, AMF in the service network, and AUSF and UDM in the core network are all wired and equipped with secure channels that can resist replay attacks, ensuring integrity and confidentiality protection \cite{33501}. In direct mode, adversaries can manipulate the wireless channel between AIoT devices and base stations, while in relay mode, adversaries can control the entire channel between AIoT devices and relays, with the relay-to-base station link secured by a channel that resists replay attacks and provides integrity and confidentiality protection \cite{33501}.

\subsection{Security Goal}

AIoT devices get access to the 5G  network directly or through relay devices. Considering the extremely limited hardware resources of these terminals, which implies even more limited resources for security, a secure authentication scheme should meet the following security goals:
\begin{itemize}
	\item Lightweight: Secure access authentication schemes for AIoT devices should involve minimal computational overhead to lower terminal expenses, encompassing hardware and energy needs. This strategy aids in averting situations where security functions remain unfinished due to terminal battery depletion.
	\item Mutual authentication: AIoT devices need to authenticate the network to prevent unauthorized network access, which would waste terminal energy and potentially lead to subsequent data leaks. When terminals transmit data to the network, the network needs to authenticate the terminals to ensure the authenticity of the data source for traffic billing purposes.
	\item Session key agreement:  AIoT devices should establish a key with the network to protect the application data.
	\item AIoT Device Authorization: For aided nodes (such as mobile phones and CPEs), which belong to individual users, they do not want to help unauthorized AIoT devices for communication. Therefore, AIoT devices need to be authorized.
	\item 	Privacy protection of identity: Avoid linkability and traceability for AIoT devices. For instance, if Alice carries an AIoT device, direct network access of the AIoT device may result in linkable and traceable attacks.
\end{itemize}

\section{The Proposed Protocols}

We introduce three secure authentication protocols, each based on SQN, nonces, and  physical layer key (PLK), respectively.  The protocols all consist of the  two phases: the registration phase and secure access phase. 

\subsection{Protocol based on SQN for Scenario1}
\subsubsection{Registration Phase}
\ 
\newline
\indent 
The  AIoT devices need to pre-subscribe to AIoT communication services in the UDM. Both the AIoT devices and the UDM store shared keys $K$ and $SQN$, along with the device's permanent identity $AIoT$ $ID$ and temporary identity $TID$. The phase is conducted securely offline.

\begin{figure*}[!htb]
\centering
  \includegraphics[width=18cm]{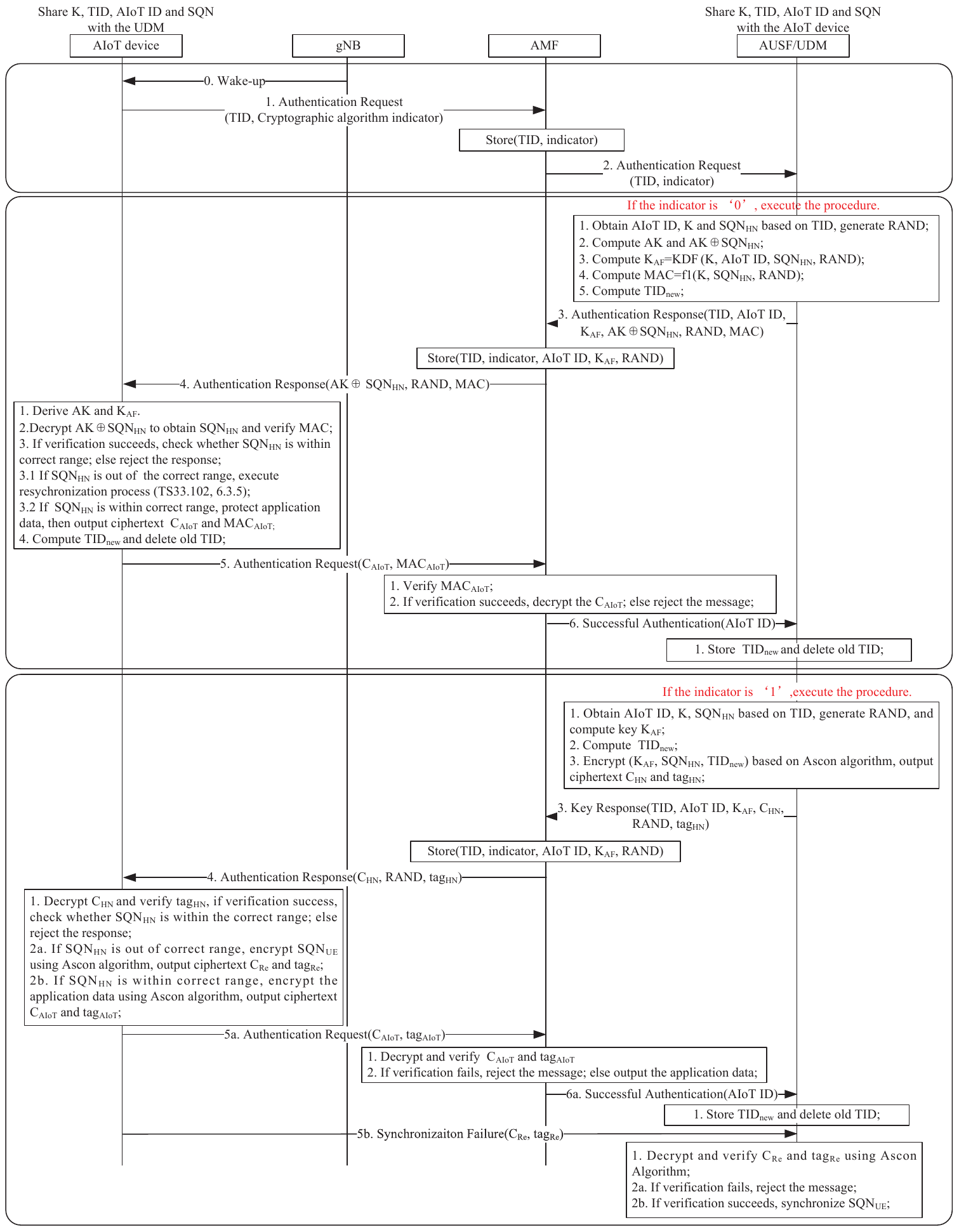}
  \caption{Protocol based on SQN for scenario1}
  \label{protocol1}
\end{figure*}
\subsubsection{Secure Access Phase}
\ 
\newline
\indent 
 As shown  in Fig \ref{protocol1}, The detailed protocol flows  are described as following.

\begin{enumerate}
  \item \textbf{gNB$\to$Device:} 
  The gNB actively sends energy carrier and trigger signals to wake up the AIoT device.
  \item \textbf{Device$\to$AMF:}
  After the AIoT device acquires energy, it transmits the authentication request message to the gNB through backscatter communication. Subsequently, the gNB forwards the message to the AMF, encompassing the $TID$ and a  cryptographic algorithm  indicator (which is only 1 bit; if the indicator is '0', it indicates the  traditional ciphering algorithms, while if the indicator is '1', it indicates the Ascon algorithm \citen{ascon}).
  \item \textbf{AMF$\to$UDM:}
  Upon receiving this message, the AMF stores the $TID$ and the  indicator, and sends an authentication request message to the  UDM associated with the device belongs through the AUSF.
  \item  \textbf{UDM$\to$AMF:}
  Upon receiving the message, the UDM retrieves the corresponding permanent identifier $AIoT$ $ID$, key $K$, and sequence number $SQN_{HN}$ based on the $TID$. The UDM then generates a random number $RAND$ and executes the subsequent procedures based on the  indicator:
  
  \textbf{If the indicator is '0':}
      \begin{enumerate}
        \item Calculate $AK$ = $f5(K, RAND)$ and then compute $AK$ $\oplus$ $SQN_{HN}$.
        \item Calculate  $K_{AF}$ = $KDF($ $K$, $AIoT$ $ID$, $SQN_{HN}$, $RAND$).
        \item Compute $MAC$ = $f1($ $K$, $SQN_{HN}$, $RAND$). 
        \item Compute new tempopary identity $TID_{new}$ = $KDF($ $K$, $AIoT$ $ID$, $TID$). 
        \item Send an authentication response message to AMF, which includes ($TID$, $AIoT$ $ID$, $K_{AF}$, $AK$ $\oplus$ $SQN_{HN}$, $RAND$, $MAC$).
      \end{enumerate}
  \textbf{If the indicator is '1':}
\begin{enumerate}
        \item Calculate  $K_{AF}$ = $KDF($ $K$, $AIoT$ $ID$, $SQN_{HN}$, $RAND$).
	\item Compute new tempopary identity $TID_{new}$ = $KDF($ $K$, $AIoT$ $ID$, $TID$). 
	\item Calculate ($C_{HN}$, $tag_{HN}$) =  $\varepsilon$ ($K$, $RAND$, $P$), $P$ = ($K_{AF}$, $SQN_{HN}$, $TID_{new}$).
	\item Send an authentication response message to AMF, which includes ($TID$, $AIoT$ $ID$, $K_{AF}$, $AK$ $\oplus$ $SQN_{HN}$, $RAND$, $MAC$).
\end{enumerate}  
  \item \textbf{AMF$\to$Device:}
  Upon receiving the authentication response message, the AMF stores together  $TID$, indicator, $AIoT$ $ID$, $K_{AF}$, and $RAND$. Then, the AMF directly forwards either ($AK$ $\oplus$ $SQN_{HN}$, $RAND$, $MAC$) or ($C_{HN}$, $RAND$, $tag_{HN}$) to the AIoT device through gNB.
  \item \textbf{Device$\to$AMF:}
  Upon receiving the  message, executes the following procedures based on the  indicator:
  
  \textbf{If the indicator is '0':}
      \begin{enumerate}
	\item Employ the same method as the network side to calculate $AK$ and $K_{AF}$.
	\item Decrypt $AK$ $\oplus$ $SQN_{HN}$ to obtain $SQN_{HN}$ and verify the received $MAC$.
	\item If verification succeeds, check whether $SQN_{HN}$ is within correct range; else reject the response.
	\item If $SQN_{HN}$ is out of  the correct range, execute resychronization process (referring to Section 6.3.5 in TS 33.102 \cite{33.102}); If $SQN_{HN}$ is within correct range, the AIoT device provides encryption and  integrity protection to application information $data$ using  $K_{AF}$ and AES algorithm, resulting in the ciphertext $C_{AIoT}$ and the authentication code $MAC_{AIoT}$. 
	\item Compute $TID_{new}$ = $KDF($ $K$, $AIoT$ $ID$, $TID$) and delete old $TID$. 
	\item Finally, the AIoT device sends ($C_{AIoT}$, $MAC_{AIoT}$) to the AMF through gNB.
\end{enumerate}

  \textbf{If the indicator is '1':}
\begin{enumerate}
	\item Calculate  $D($ $K$, $RAND$, $C_{HN}$, $tag_{HN}$), if the result is $\perp$, reject the response; otherelse, if the result is ($K_{AF}$, $SQN_{HN}$, $TID_{new}$), check whether $SQN_{HN}$ is within correct range.
	\item  If $SQN_{HN}$ is out of correct range, calculate ($C_{Re}$, $tag_{Re}$) =  $\varepsilon$ ($K_{AF}$, $RAND$, $data$);
	finally, send  ($C_{Re}$, $tag_{Re}$) to the AMF.
	\item If $SQN_{HN}$  is within correct range, calculate ($C_{AIoT}$, $tag_{AIoT}$) =  $\varepsilon$ ($K_{AF}$, $RAND$, $data$); store $TID_{new}$ and delete old $TID$, finally, send  ($C_{AIoT}$, $tag_{AIoT}$) to the AMF.
\end{enumerate} 

  \item \textbf{AMF$\to$UDM:}
     Upon receiving the  message, executes the following procedures based on the  indicator:
     
  \textbf{If the indicator is '0':}
\begin{enumerate}
	\item Verify  $MAC_{AIoT}$ using $K_{AF}$.
	\item If verification succeeds, decrypt  $C_{AIoT}$; else  reject the message.
	\item Send authentication success message to AUSF and UDM, which includes  $AIoT$ $ID$.
\end{enumerate}

  \textbf{If the indicator is '1':}
\begin{enumerate}
	\item Calculate  $D$ = ($K_{AF}$, $RAND$, $C_{AIoT}$, $tag_{AIoT}$).
	\item If the result is $\perp$, reject the message; else output $data$.
	\item Send authentication success message to AUSF and UDM, which includes  $AIoT$ $ID$.
\end{enumerate}
     \item Upon receiving the authentication success message, UDM stores $TID_{new}$ corresponding to $AIoT$ $ID$ and deletes the old $TID$. If UDM receives a synchronization failure message ($C_{Re}$, $tag_{Re}$), it decrypts and verifies the ciphertext $C_{Re}$ and $tag_{Re}$ using the Ascon algorithm. If the verification of $tag_{Re}$ fails, the message is rejected. If the verification of $tag_{Re}$ succeeds, $SQN_{UE}$ is synchronized.
\end{enumerate}

\subsection{Protocol based on SQN for Scenario2}

In this scenario, the registration stage is the same as in scenario1. The difference in the secure access stage lies in the following:  The third-party application server  sends a  request message to the UDM through the NEF. This message includes the $AIoT$ $ID$, indicator, and an optional AMF ID (because UDM can choose AMF based on the received AMF ID or retrieve the previous AMF ID using the $AIoT$ $ID$, as described in Section  5.2.3.2.4 in TS 23.502 \cite{23502}). Upon receiving this request message, the UDM proceeds with the same process as in scenario1.

\subsection{Protocol based on SQN for Scenario3}

\subsubsection{Registration Phase}
\ 
\newline
\indent 
The registration phase of this scenario encompasses identical elements to those of scenario1, with additional aspects: the AIoT device shares its $AIoT$ $ID$ and $TID$ with the UE, which manages  the mapping among $AIoT$ $ID$, $TID$, and $SUPI$. Furthermore, the UDM shares this mapping with the UE.

\subsubsection{Secure Access Phase}
\ 
\newline
\indent 
The difference between this phase and the one in scenario1 is as follows: UE replaces gNB in scenario1. Additionally, upon receiving the authentication request message, UE checks whether there is a mapping between $TID$ and $SUPI$. If there is no mapping, the request is rejected. If there is a mapping, then UE sends an authentication request message to the AMF through gNB.
After receiving the authentication success message, UDM sends $TID_{new}$ to UE. Upon receiving the message, UE deletes  old $TID$ and stores $TID_{new}$.

\subsection{Protocol based on SQN for Scenario4}

The registration phase is the same as in scenario3. The difference in the access phase compared to scenario3 is as follows:  The third-party application server  sends a  request message to UDM through NEF. This message includes $UE$ $ID$, $AIoT$ $ID$, and indicator. The subsequent process is identical to the secure access phase in scenario3.

\subsection{Protocol based on nonce for Scenario1}
\subsubsection{Registration Phase}
\ 
\newline
\indent 
The  AIoT devices need to pre-subscribe to AIoT communication services in the UDM. Both the AIoT devices and the UDM store shared keys $K$, along with the device's permanent identity $AIoT$ $ID$ and temporary identity $TID$. The phase is conducted securely offline.

\begin{figure*}[!htb]
	\centering
	\includegraphics[width=18cm]{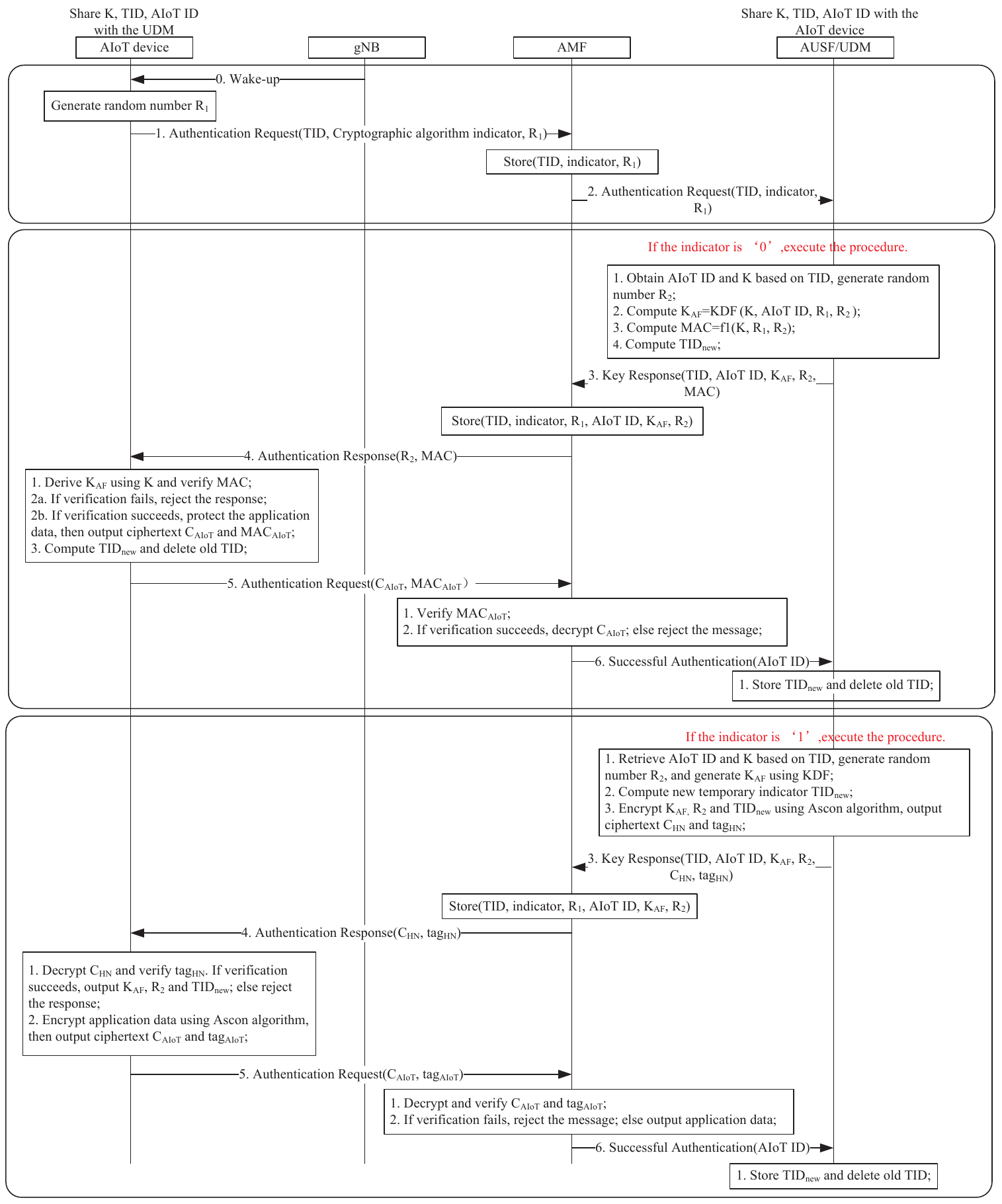}
	\caption{Protocol based on nonce for scenario1}
	\label{protocol2}
\end{figure*}

\subsubsection{Secure Access Phase}
\ 
\newline
\indent 
As shown  in Fig \ref{protocol2}, The detailed protocol flows  are described as following.

\begin{enumerate}
	\item \textbf{gNB$\to$Device:} 
	The gNB actively sends energy carrier and trigger signals to wake up the AIoT device.
	\item \textbf{Device$\to$AMF:}
	After the AIoT device acquires energy, it transmits the authentication request message to the gNB through backscatter communication. Subsequently, the gNB forwards the message to the AMF, encompassing the $TID$, $R_1$ (the device can pre-generate the nonce) and a  cryptographic algorithm  indicator (same as the one in protocol based on SQN).
	\item \textbf{AMF$\to$UDM:}
	Upon receiving this message, the AMF stores the $TID$,  the  indicator and $R_1$, and sends an authentication request message to the  UDM associated with the device belongs through the AUSF.
	\item  \textbf{UDM$\to$AMF:}
	Upon receiving the message, the UDM retrieves the corresponding $AIoT$ $ID$, $K$ based on the $TID$. The UDM then generates a random number $R_2$ and executes the subsequent procedures based on the  indicator:
	
	\textbf{If the indicator is '0':}
	\begin{enumerate}
		\item Calculate  $K_{AF}$ = $KDF($ $K$, $AIoT$ $ID$, $R_1$, $R_2$).
		\item Compute $MAC$ = $f1($ $K$, $R_1$, $R_2$). 
		\item Compute $TID_{new}$ = $KDF($ $K$, $AIoT$ $ID$, $TID$). 
		\item Send an authentication response message to AMF, which includes ($TID$, $AIoT$ $ID$, $K_{AF}$, $R_2$, $MAC$).
	\end{enumerate}
	\textbf{If the indicator is '1':}
	\begin{enumerate}
		\item Calculate  $K_{AF}$ = $KDF($ $K$, $AIoT$ $ID$, $R_1$, $R_2$).
		\item Compute  $TID_{new}$ = $KDF($ $K$, $AIoT$ $ID$, $TID$). 
		\item Calculate ($C_{HN}$, $tag_{HN}$) =  $\varepsilon$ ($K$, $R_1$, $P$), $P$ = ($K_{AF}$, $R_2$, $TID_{new}$).
		\item Send an authentication response message to AMF, which includes ($TID$, $AIoT$ $ID$, $K_{AF}$, $R_2$, $C_{HN}$, $tag_{HN}$).
	\end{enumerate}  
	\item \textbf{AMF$\to$Device:}
	Upon receiving the authentication response message, the AMF stores together  $TID$, indicator, $R_1$, $AIoT$ $ID$, $K_{AF}$, and $R_2$. Then, the AMF directly forwards either ($R_2$, $MAC$) or ($C_{HN}$, $tag_{HN}$) to the AIoT device through gNB.
	\item \textbf{Device$\to$AMF:}
	Upon receiving the  message, executes the following procedures based on the  indicator:
	
	\textbf{If the indicator is '0':}
	\begin{enumerate}
		\item Employ the same method as the network side to calculate $K_{AF}$ and verify the received $MAC$..
		\item If verification fails, reject the response.
		\item If verification succeeds, the AIoT device provides encryption and  integrity protection to application information $data$ using  $K_{AF}$ and AES algorithm, resulting in the ciphertext $C_{AIoT}$ and the authentication code $MAC_{AIoT}$. 
		\item Compute $TID_{new}$ = $KDF($ $K$, $AIoT$ $ID$, $TID$) and delete old $TID$. 
		\item Finally, the AIoT device sends ($C_{AIoT}$, $MAC_{AIoT}$) to the AMF through gNB.
	\end{enumerate}
	
	\textbf{If the indicator is '1':}
	\begin{enumerate}
		\item Calculate  $D($ $K$, $R_1$, $C_{HN}$, $tag_{HN}$), if the result is $\perp$, reject the response; otherelse, output  ($K_{AF}$, $R_2$, $TID_{new}$).
		\item Calculate ($C_{AIoT}$, $tag_{AIoT}$) =  $\varepsilon$ ($K_{AF}$, $R_2$, $data$); store $TID_{new}$ and delete old $TID$, finally, send  ($C_{AIoT}$, $tag_{AIoT}$) to the AMF.
	\end{enumerate} 
	
	\item \textbf{AMF$\to$UDM:}
	Upon receiving the  message, executes the following procedures based on the  indicator:
	
	\textbf{If the indicator is '0':}
	\begin{enumerate}
		\item Verify  $MAC_{AIoT}$ using $K_{AF}$.
		\item If verification succeeds, decrypt  $C_{AIoT}$; else  reject the message.
		\item Send authentication success message to AUSF and UDM, which includes  $AIoT$ $ID$.
	\end{enumerate}
	
	\textbf{If the indicator is '1':}
	\begin{enumerate}
		\item Calculate  $D$ = ($K_{AF}$, $RAND$, $C_{AIoT}$, $tag_{AIoT}$).
		\item If the result is $\perp$, reject the message; else output $data$.
		\item Send authentication success message to AUSF and UDM, which includes  $AIoT$ $ID$.
	\end{enumerate}
	\item Upon receiving the authentication success message, UDM stores $TID_{new}$ corresponding to $AIoT$ $ID$ and deletes the old $TID$.
\end{enumerate}

\subsection{Protocol based on nonce for Scenario2}

The scenario  has the same registration and access phases as Scenario1. The difference lies in the access phase, where the third-party application server  needs to actively send a request message to activate the AIoT terminal through gNB, thereby establishing secure communication between the network and a specific AIoT terminal.

\subsection{Protocol based on nonce for Scenario3}

\subsubsection{Registration Phase}
\ 
\newline
\indent 
The registration phase of this scenario encompasses identical elements to those of Scenario1, with additional aspects: the AIoT device shares its $AIoT$ $ID$ and $TID$ with the UE, which manages  the mapping among $AIoT$ $ID$, $TID$, and $SUPI$. Furthermore, the UDM shares this mapping with the UE.

\subsubsection{Secure Access Phase}
\ 
\newline
\indent 
The difference between this phase and the one in Scenario1 is as follows: UE replaces gNB in Scenario1. Additionally, upon receiving the authentication request message, UE checks whether there is a mapping between $TID$ and $SUPI$. If there is no mapping, the request is rejected. If there is a mapping, then UE sends an authentication request message to the AMF through gNB.
After receiving the authentication success message, UDM sends $TID_{new}$ to UE. Upon receiving the message, UE deletes  old $TID$ and stores $TID_{new}$.

\subsection{Protocol based on nonce for Scenario4}

The registration phase is the same as in Scenario3. The difference in the access phase compared to Scenario3 is as follows:  The third-party application server  sends a  request message to UDM through NEF. This message includes $UE$ $ID$, $AIoT$ $ID$, and indicator.  The network activates the AIoT device through gNB or UE, the subsequent process is identical to the secure access phase in Scenario3.

\subsection{Protocol based on PLK for Scenario1}
\subsubsection{Registration Phase}
\ 
\newline
\indent 
The  phase is same as the one in the protocol based on nonce for scenario1.

\begin{figure*}[!htb]
	\centering
	\includegraphics[width=18cm]{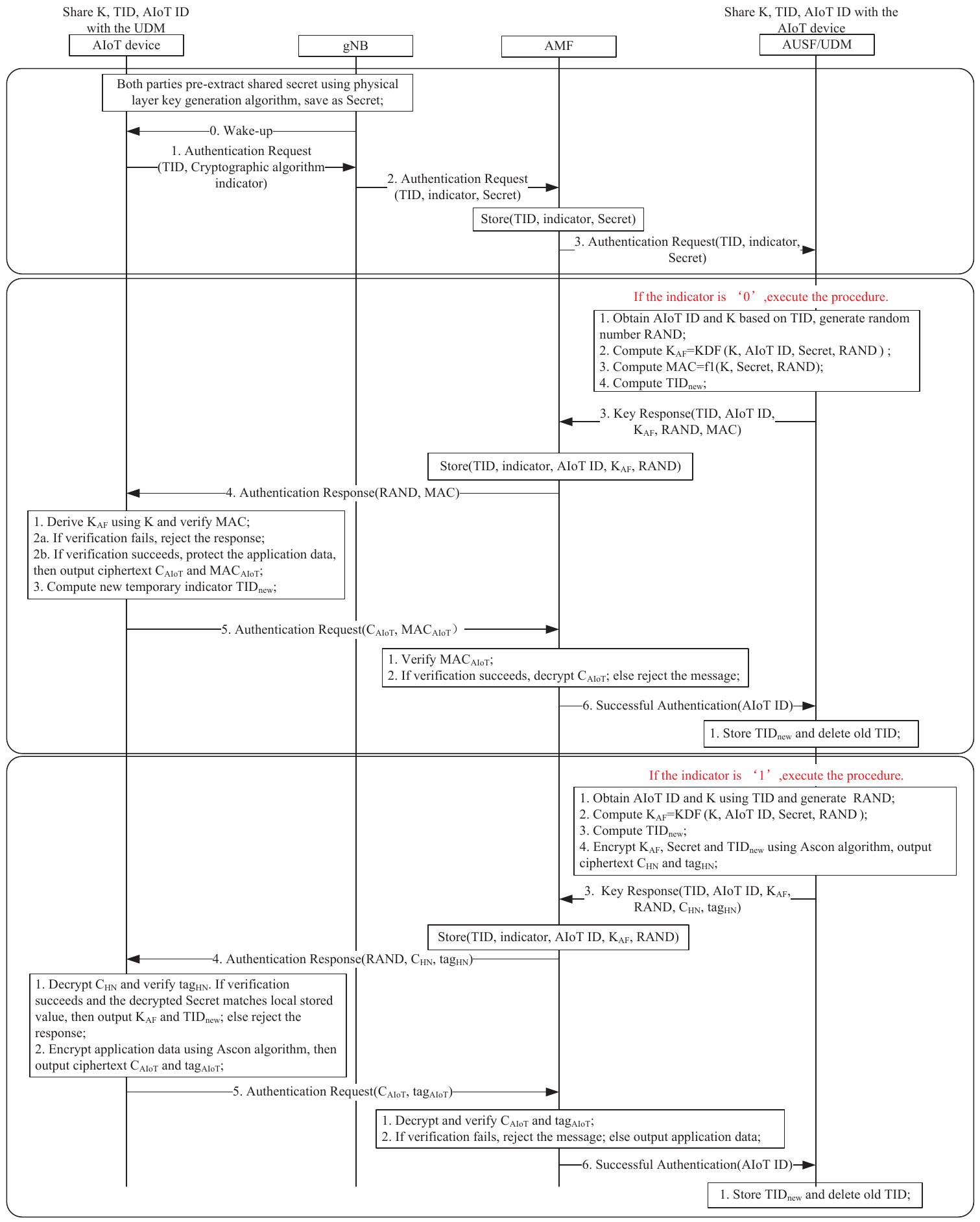}
	\caption{Protocol based on PLK for scenario1}
	\label{protocol3}
\end{figure*}

\subsubsection{Secure Access Phase}
\ 
\newline
\indent 
As shown  in Fig \ref{protocol3}, The detailed protocol flows  are described as following.

\begin{enumerate}
	\item \textbf{gNB$\to$Device:} 
	The AIoT device and the gNB pre-extract a shared secret using a physical layer key generation algorithm, denoted as Secret. The gNB actively sends energy carrier and trigger signals to wake up the AIoT device.
	\item \textbf{Device$\to$AMF:}
	After the AIoT device acquires energy, it transmits  $TID$, and a   indicator (same as the one in protocol based on SQN) to the gNB through backscatter communication. Then the gNB forwards ($TID$, indicator, $Secret$) to the AMF, 
	\item \textbf{AMF$\to$UDM:}
	Upon receiving this message, the AMF stores the $TID$,  the  indicator and $Secret$, and sends an authentication request message to the  UDM associated with the device belongs through the AUSF.
	\item  \textbf{UDM$\to$AMF:}
	Upon receiving the message, the UDM retrieves the corresponding $AIoT$ $ID$, $K$ based on the $TID$. The UDM then generates a random number $RAND$ and executes the subsequent procedures based on the  indicator:
	
	\textbf{If the indicator is '0':}
	\begin{enumerate}
		\item Calculate  $K_{AF}$ = $KDF($ $K$, $AIoT$ $ID$, $Secret$, $RAND$).
		\item Compute $MAC$ = $f1($ $K$, $Secret$, $RAND$). 
		\item Compute $TID_{new}$ = $KDF($ $K$, $AIoT$ $ID$, $TID$). 
		\item Send an authentication response message to AMF, which includes ($TID$, $AIoT$ $ID$, $K_{AF}$, $RAND$, $MAC$).
	\end{enumerate}
	\textbf{If the indicator is '1':}
	\begin{enumerate}
		\item Calculate  $K_{AF}$ = $KDF($ $K$, $AIoT$ $ID$, $Secret$, $RAND$).
		\item Compute  $TID_{new}$ = $KDF($ $K$, $AIoT$ $ID$, $TID$). 
		\item Calculate ($C_{HN}$, $tag_{HN}$) =  $\varepsilon$ ($K$, $RAND$, $P$), $P$ = ($K_{AF}$, $Secret$, $TID_{new}$).
		\item Send an authentication response message to AMF, which includes ($TID$, $AIoT$ $ID$, $K_{AF}$, $RAND$, $C_{HN}$, $tag_{HN}$).
	\end{enumerate}  
	\item \textbf{AMF$\to$Device:}
	Upon receiving the authentication response message, the AMF stores together  $TID$, indicator, $Secret$, $AIoT$ $ID$, $K_{AF}$, and $RAND$. Then, the AMF directly forwards either ($RAND$, $MAC$) or ($C_{HN}$, $tag_{HN}$) to the AIoT device through gNB.
	\item \textbf{Device$\to$AMF:}
	Upon receiving the  message, executes the following procedures based on the  indicator:
	
	\textbf{If the indicator is '0', }the procedures are same as the one in the protocol based on nonce.
	
	\textbf{If the indicator is '1':}
	\begin{enumerate}
		\item Calculate  $D($ $K$, $RAND$, $C_{HN}$, $tag_{HN}$), if verification succeeds and the decrypted $Secret$ matches local stored value, then output $K_{AF}$ and $TID_{new}$; else reject the response;
		\item Calculate ($C_{AIoT}$, $tag_{AIoT}$) =  $\varepsilon$ ($K_{AF}$, $RAND$, $data$); store $TID_{new}$ and delete old $TID$, finally, send  ($C_{AIoT}$, $tag_{AIoT}$) to the AMF.
	\end{enumerate} 
	
	\item \textbf{AMF$\to$UDM:}
	Upon receiving the  message, executes the following procedures based on the  indicator:
	
	\textbf{If the indicator is '0',} the procedures are same as the one in the protocol based on nonce.

	\textbf{If the indicator is '1',} the procedures are same as the one in the protocol based on nonce.

	\item Upon receiving the authentication success message, UDM stores $TID_{new}$ corresponding to $AIoT$ $ID$ and deletes the old $TID$.
\end{enumerate}

\subsection{Protocol based on PLK for Scenario2}
The procedures are same as the one in the protocol based on SQN for scenario2.

\subsection{Protocol based on PLK for Scenario3}
The procedures are same as the one in the protocol based on SQN for scenario3.

\subsection{Protocol based on PLK for Scenario4}
The procedures are same as the one in the protocol based on SQN for scenario4.

\section{Security Evaluation}

If not specified otherwise, the security analysis below applies to all protocols, i.e., all protocols based on SQN, nonce, and  PLK. Since the authentication processes in the four scenarios mentioned above are very similar, the informal security analysis process will only be illustrated using scenario1 as an example. The security analysis process for other scenarios is similar. Our proposed protocols can achieve the following security properties.

\textbf{Mutual Authentication and Key Agreement:} 
The AIoT device completes the authentication with the network side by verifying the received $MAC$ or $tag_{HN}$ (depending on the used cryptographic algorithm). Since the $MAC$ or $tag_{HN}$ is computed from the secret key $K$, which is shared only between the AIoT device and the network side, the AIoT device can accomplish the authentication with the network side. On the other hand, the network side performs the authentication with the AIoT device by verifying the received $MAC_{AIoT}$ or $tag_{AIoT}$. The input parameters deriving $MAC_{AIoT}$ or $tag_{AIoT}$ include $K_{AF}$,  furthermore, the input parameters deriving $K_{AF}$ include the secret key $K$, allowing the network side to authenticate the AIoT device.

According to the protocol procedure, after the protocol execution, the AIoT device and the network side will derive the key $K_{AF}$, which is used for subsequent encryption and integrity protection of application data.

\textbf{Replay Attack Resistance:} 
If an adversary replays the initial message sent by the AIoT device, the network side will respond with the corresponding authentication response message according to the protocol flow. Furthermore, if the adversary continues to replay historical messages to respond to the network side, they will be unable to pass the network side's verification because the $MAC_{AIoT}$ or $tag_{AIoT}$ in the historical message is bound to the historical random number generated by the network side (specifically, the $K_{AF}$ used to generate $MAC_{AIoT}$ or $tag_{AIoT}$ is bound to the historical random number). Additionally, since the network side generates new random numbers with each authentication protocol execution, the historical message replayed by the adversary cannot pass the verification on the network side.

If an adversary attempts to replay the second message from the network side to the AIoT device, in the SQN-based authentication protocol, the AIoT device will decrypt and obtain $SQN_{HN}$ from the replayed message, and it will find that $SQN_{HN}$  is less than the locally stored $SQN_{UE}$. In the nonce-based authentication protocol, since the $MAC$ in the replayed message is bound with historical $R_1$, and the AIoT device generates a new $R_1$ with each authentication process, the AIoT device can determine the validity of the replayed message based on the new $R_1$ in this authentication process. As for the authentication protocol based on PLK, since the $MAC$ or $tag_{HN}$ in the response returned by the network side is bound with the shared $Secret$ generated by the physical layer key generation algorithm, and the $Secret$ is only shared between the device and the network side, and both parties update this value before each authentication process, the AIoT device can determine whether the received message is a historical replay message based on this value.

\textbf{Man-in-the-Middle Attack Resistance:}
 The attack would lead to the following consequences: the AIoT device mistakenly believes it is communicating with the network side, while the network side mistakenly believes it is communicating with the AIoT device. In reality, they are both communicating with the adversary, enabling the adversary to eavesdrop on and tamper with their communication content. To successfully execute such an attack in the aforementioned protocols, the adversary needs to forge $MAC_{AIoT}$ or $tag_{AIoT}$ corresponding to the legitimate AIoT device, as well as forge  $MAC$ or $tag_{HN}$  corresponding to the network side. Based on the security  analysis of mutual authentication property, the protocol demonstrates the capability to resist man-in-the-middle attacks.
 
\textbf{Impersonation Attack Resistance:}
If an adversary attempts to forge messages to impersonate a legitimate AIoT device, according to the security analysis of mutual authentication properties, the adversary cannot forge $MAC_{AIoT}$ or $tag_{AIoT}$ corresponding to a legitimate AIoT device, nor can they forge $MAC$ or $tag_{HN}$ corresponding to the network side. Therefore, the protocol has the capability to resist impersonation attacks.

\textbf{Identity Privacy Protection:} 
We do not use existing public key technology  to protect identity privacy. Instead, the AIoT device employs a temporary identity  $TID$ to avoid exposure of the true permanent identity. Moreover, upon completion of the protocol execution, both the AIoT device and the network side generate a new temporary identity $TID_{new}$ for the next session. Since the input parameters deriving $TID_{new}$ include the $K$ and $TID_{new}$ is not transmitted over public channels, confidentiality of the temporary identity is achieved. Consequently, the unlinkability and untraceability of  the AIoT device is ensured.

\textbf{AIoT Device Authorization:}
The security requirement only arises in scenario3 and scenario4. In the protocols designed for scenario3 and scenario4, the UE can authorize the AIoT device based on its temporary identity. If the adversary uses an illegal temporary identity, the UE will reject the adversary's request because it does not maintain a mapping relationship with the illegal identity.

\section{Performance Analysis}

In this section, we will evaluate the computational cost and energy cost among the proposed protocols with other related protocols.

\subsection{Computational Overhead}

\begin{figure}[!htb]
	\centering
	\includegraphics[width=9cm]{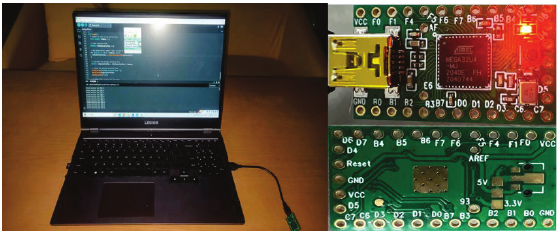}
	\caption{Test environment for protocols}
	\label{test}
\end{figure}

Considering the fact that the network side possesses  very powerful computing resources and different protocols have minimal impact on the network side in terms of computational overhead, but the situation is opposite on the device side, therefore, we will only analyze the computational overhead on the device side.
We will compare the proposed protocols with others in the current standard in terms of computational cost. Since the running time of the proposed protocols mainly depends on the involved cryptographic operations, we will only calculate the executation overhead of the relevant cryptographic operations. 
The processing capability for a AIoT device is likely to be 4MHz by contacting members of the 3GPP working group.
We use Teensy 2.0 Development Board with an ATMEGA32U4 processor (8 bit, AVR16 MHz) \cite{test1} and Arduino Cryptography Library \cite{test2} to evaluate these cryptographic operations, and we redifined the clock frequency to 4MHz to simulate the AIoT devices and modified the \textit{micros()}  fucntion to adapt to the new frequency. The test environment can be seen in Fig. \ref{test}.

As all f1/f2/f3/f4/f5 algorithms  are executed by invoking AES-ECB algorithm, and its input size and key size both can be 128bits, we only test the corresponding AES-ECB algorithm. There are three different encryption/integrity protection algorithms in 3GPP standard for data transmission, we take AES-CBC for encryption and AES-CMAC for integrity protection as an example, and we set the key size as 128bits, the application data as 128bits.
For 5G AKA, the computational overhead of cryptographic operations  are listed in Table \ref{5gaka_cost_overhead}.
For EAP-AKA', the computational overhead of cryptographic operations  are listed in Table \ref{eapaka_cost_overhead}.
For EPS-AKA, the computational overhead of cryptographic operations  are listed in Table \ref{epsaka_cost_overhead}.
For CP CIoT AKA, the computational overhead of cryptographic operations  are listed in Table \ref{CIoT_cost_overhead}.
For BEST AKA based on 5G AKA (denoted as BEST AKA1), the computational overhead of cryptographic operations  are listed in Table \ref{BEST1_cost_overhead}.
For BEST AKA based on EAP-AKA' (denoted as BEST AKA2), the computational overhead of cryptographic operations  are listed in Table \ref{BEST2_cost_overhead}.
For BEST AKA based on EPS AKA (denoted as BEST AKA3), the computational overhead of cryptographic operations  are listed in Table \ref{BEST3_cost_overhead}.
For proposed AKA based on SQN, the computational overhead of cryptographic operations  are listed in Table \ref{SQN1_cost_overhead} and \ref{SQN2_cost_overhead}.
Since the nonce and physical layer shared key can be generated at idle time for AIoT device, we ignore the computation time here.
For proposed AKA based on nonce or PLK, the computational overhead of cryptographic operations  are same and listed in Table \ref{nonce1_cost_overhead} and \ref{nonce2_cost_overhead}.

The comparison results about the computational overhead among all protocols can be seen in Table \ref{cost_overhead}. It can be seen that the   execution time for all existing protocols to complete authentication and one round of data transmission is at least 2 seconds. This complexity arises from the intricate key hierarchy in existing protocols, significantly diminishing the efficiency of AIoT devices when accessing the network and imposing pressure on the devices, such as if the AIoT device is battery-free, it takes a long time to obtain energy, or if the AIoT device is equipped with a capacitor, the battery may be exhausted before the protocol is completed. By contrast, all our proposed schemes are less than 1s, especially the Ascon-based protocol has the smallest running time. This is due to the fact that we simplified the key hierarchy  and we integrate secure transmission and authentication process (i.e., when the AIoT device responses its application data, the network side can authenticate the AIoT device), which results in the reduction of the number of the HMAC operations directly and significant reduction in time. 

\begin{table}[!htb]
	\renewcommand{\arraystretch}{1.3}
	\caption{computational cost for 5G AKA}
	\label{5gaka_cost_overhead}
	\centering
	\resizebox{8.8cm}{!}{
	\begin{tabular}{|c|c|c|c|}
		\hline
		Operations & Key Size(bits) & Input Size(bits) & time(us) \\
		\hline
		\multicolumn{4}{|c|}{AKA Phase}\\
		\hline
		fi(CK,IK,AK,MAC,XRES)  & 128 & 128 & 3515*5\\
		\hline
		HMAC$^1$$\to$RES* & 256 & 480 & 352417\\
		\hline
		HMAC$\to$$K_{AUSF}$ & 256 & 328 & 352417\\
		\hline
		HMAC$\to$$K_{SEAF}$ & 256 & 272 & 352417\\
		\hline
		HMAC$\to$$K_{AMF}$ & 256 & 104 & 352417\\
		\hline
		HMAC$\to$$K_{NAS-enc}$ & 256 & 40 & 352417\\
		\hline
		HMAC$\to$$K_{NAS-int}$ & 256 & 40 & 352417\\
		\hline
		HMAC$\to$$K_{RRC-enc}$ & 256 & 40 & 352417\\
		\hline
		HMAC$\to$$K_{RRC-int}$ & 256 & 40 & 352417\\
		\hline
		HMAC$\to$$K_{UP-enc}$ & 256 & 40 & 352417\\
		\hline
		HMAC$\to$$K_{UP-int}$ & 256 & 40 & 352417\\					
		\hline
		\multicolumn{4}{|c|}{Data Transmission Phase}\\
		\hline
		AES-CBC/CMAC  & 128 & 128 & 3536+6219\\	
		\hline
	\end{tabular}}
\begin{tablenotes}
	\item[1] 1.Although the input size or key size for  HMAC operation is different, the effect  was actually found to be negligible. So we set the operation time is same for easy to calculate.
\end{tablenotes}
\end{table}

\begin{table}[!htb]
	\renewcommand{\arraystretch}{1.3}
	\caption{computational cost for EAP-AKA'}
	\label{eapaka_cost_overhead}
	\centering
	\resizebox{8.8cm}{!}{
		\begin{tabular}{|c|c|c|c|}
			\hline
			Operations & Key Size(bits) & Input Size(bits) & time(us) \\
			\hline
			\multicolumn{4}{|c|}{AKA Phase}\\
			\hline
			fi(CK,IK,AK,MAC,XRES)  & 128 & 128 & 3515*5\\
			\hline
			HMAC$\to$$AT\_{MAC}\_HN$ & 256 & 576 & 352417\\
			\hline
			HMAC$\to$$AT\_{MAC}\_UE$ & 256 & 64 & 352417\\
			\hline
			HMAC$\to$$CK'||IK'$ & 256 & 96 & 352417\\
			\hline
			HMAC$\to$$K_{AUSF}^1$ & 256 & 408 & 352417\\
			\hline
			HMAC$\to$$K_{SEAF}$ & 256 & 272 & 352417\\
			\hline
			HMAC$\to$$K_{AMF}$ & 256 & 104 & 352417\\
			\hline
			HMAC$\to$$K_{NAS-enc}$ & 256 & 40 & 352417\\
			\hline
			HMAC$\to$$K_{NAS-int}$ & 256 & 40 & 352417\\
			\hline
			HMAC$\to$$K_{RRC-enc}$ & 256 & 40 & 352417\\
			\hline
			HMAC$\to$$K_{RRC-int}$ & 256 & 40 & 352417\\
			\hline
			HMAC$\to$$K_{UP-enc}$ & 256 & 40 & 352417\\
			\hline
			HMAC$\to$$K_{UP-int}$ & 256 & 40 & 352417\\					
			\hline
			\multicolumn{4}{|c|}{Data Transmission Phase}\\
			\hline
			AES-CBC/CMAC  & 128 & 128 & 3536+6219\\	
			\hline
	\end{tabular}}
\begin{tablenotes}
	\item[1] 1.It's noted that it needs to execute four HMAC-sha256 operations to derive $K_{AUSF}$. 
\end{tablenotes}
\end{table}

\begin{table}[!htb]
	\renewcommand{\arraystretch}{1.3}
	\caption{computational cost for EPS-AKA}
	\label{epsaka_cost_overhead}
	\centering
	\resizebox{8.8cm}{!}{
		\begin{tabular}{|c|c|c|c|}
			\hline
			Operations & Key Size(bits) & Input Size(bits) & time(us) \\
			\hline
			\multicolumn{4}{|c|}{AKA Phase}\\
			\hline
			fi(CK,IK,AK,MAC,XRES)  & 128 & 128 & 3515*5\\
			\hline
			HMAC$\to$$K_{ASME}$ & 256 & 96 & 352417\\
			\hline
			HMAC$\to$$K_{NAS-enc}$ & 256 & 40 & 352417\\
			\hline
			HMAC$\to$$K_{NAS-int}$ & 256 & 40 & 352417\\
			\hline
			HMAC$\to$$K_{eNB}$ & 256 & 48 & 352417\\
			\hline
			HMAC$\to$$K_{RRC-enc}$ & 256 & 40 & 352417\\
			\hline
			HMAC$\to$$K_{RRC-int}$ & 256 & 40 & 352417\\
			\hline
			HMAC$\to$$K_{UP-enc}$ & 256 & 40 & 352417\\
			\hline
			HMAC$\to$$K_{UP-int}$ & 256 & 40 & 352417\\					
			\hline
			\multicolumn{4}{|c|}{Data Transmission Phase}\\
			\hline
			AES-CBC/CMAC  & 128 & 128 & 3536+6219\\	
			\hline
	\end{tabular}}
\end{table}

\begin{table}[!htb]
	\renewcommand{\arraystretch}{1.3}
	\caption{computational cost for CP CIoT AKA}
	\label{CIoT_cost_overhead}
	\centering
	\resizebox{8.8cm}{!}{
		\begin{tabular}{|c|c|c|c|}
			\hline
			Operations & Key Size(bits) & Input Size(bits) & time(us) \\
			\hline
			\multicolumn{4}{|c|}{AKA Phase}\\
			\hline
			fi(CK,IK,AK,MAC,XRES)  & 128 & 128 & 3515*5\\
			\hline
			HMAC$\to$RES* & 256 & 480 & 352417\\
			\hline
			HMAC$\to$$K_{AUSF}$ & 256 & 328 & 352417\\
			\hline
			HMAC$\to$$K_{SEAF}$ & 256 & 272 & 352417\\
			\hline
			HMAC$\to$$K_{AMF}$ & 256 & 104 & 352417\\
			\hline
			HMAC$\to$$K_{NAS-enc}$ & 256 & 40 & 352417\\
			\hline
			HMAC$\to$$K_{NAS-int}$ & 256 & 40 & 352417\\			
			\hline
			\multicolumn{4}{|c|}{Data Transmission Phase}\\
			\hline
			AES-CBC/CMAC  & 128 & 128 & 3536+6219\\	
			\hline
	\end{tabular}}
\end{table}

\begin{table}[!htb]
	\renewcommand{\arraystretch}{1.3}
	\caption{computational cost for BEST AKA1}
	\label{BEST1_cost_overhead}
	\centering
	\resizebox{8.8cm}{!}{
		\begin{tabular}{|c|c|c|c|}
			\hline
			Operations & Key Size(bits) & Input Size(bits) & time(us) \\
			\hline
			\multicolumn{4}{|c|}{AKA Phase}\\
			\hline
			fi(CK,IK,AK,MAC,XRES)  & 128 & 128 & 3515*5\\
			\hline
			HMAC$\to$RES* & 256 & 480 & 352417\\
			\hline
			HMAC$\to$$K_{HSE}$ & 256 & 328 & 352417\\
			\hline
			HMAC$\to$$K_{E2M-enc}$ & 256 & 112 & 352417\\
			\hline
			HMAC$\to$$K_{E2M-int}$ & 256 & 112 & 352417\\
			\hline
			HMAC$\to$$K_{intermediate}$ & 256 & 112 & 352417\\
			\hline
			HMAC$\to$$K_{EAS-PSK}$ & 256 & 40 & 352417\\
			\hline
			HMAC$\to$$K_{E2E-enc}$ & 256 & 24 & 352417\\
			\hline
			HMAC$\to$$K_{E2E-int}$ & 256 & 24 & 352417\\			
			\hline
			\multicolumn{4}{|c|}{Data Transmission Phase}\\
			\hline
			AES-CBC/CMAC  & 128 & 128 & 3536+6219\\	
			\hline
	\end{tabular}}
\end{table}

\begin{table}[!htb]
	\renewcommand{\arraystretch}{1.3}
	\caption{computational cost for BEST AKA2}
	\label{BEST2_cost_overhead}
	\centering
	\resizebox{8.8cm}{!}{
		\begin{tabular}{|c|c|c|c|}
			\hline
			Operations & Key Size(bits) & Input Size(bits) & time(us) \\
			\hline
			\multicolumn{4}{|c|}{AKA Phase}\\
			\hline
			fi(CK,IK,AK,MAC,XRES)  & 128 & 128 & 3515*5\\
			\hline
			HMAC$\to$RES* & 256 & 480 & 352417\\
			\hline
			HMAC$\to$$CK'||IK'$ & 256 & 96 & 352417\\
			\hline
			HMAC$\to$$K_{HSE}$ & 256 & 328 & 352417\\
			\hline
			HMAC$\to$$K_{E2M-enc}$ & 256 & 112 & 352417\\
			\hline
			HMAC$\to$$K_{E2M-int}$ & 256 & 112 & 352417\\
			\hline
			HMAC$\to$$K_{intermediate}$ & 256 & 112 & 352417\\
			\hline
			HMAC$\to$$K_{EAS-PSK}$ & 256 & 40 & 352417\\
			\hline
			HMAC$\to$$K_{E2E-enc}$ & 256 & 24 & 352417\\
			\hline
			HMAC$\to$$K_{E2E-int}$ & 256 & 24 & 352417\\			
			\hline
			\multicolumn{4}{|c|}{Data Transmission Phase}\\
			\hline
			AES-CBC/CMAC  & 128 & 128 & 3536+6219\\	
			\hline
	\end{tabular}}
\end{table}

\begin{table}[!htb]
	\renewcommand{\arraystretch}{1.3}
	\caption{computational cost for BEST AKA3}
	\label{BEST3_cost_overhead}
	\centering
	\resizebox{8.8cm}{!}{
		\begin{tabular}{|c|c|c|c|}
			\hline
			Operations & Key Size(bits) & Input Size(bits) & time(us) \\
			\hline
			\multicolumn{4}{|c|}{AKA Phase}\\
			\hline
			fi(CK,IK,AK,MAC,XRES)  & 128 & 128 & 3515*5\\
			\hline
			HMAC$\to$$K_{ASME}$ & 256 & 96 & 352417\\
			\hline
			HMAC$\to$$K_{E2M-enc}$ & 256 & 112 & 352417\\
			\hline
			HMAC$\to$$K_{E2M-int}$ & 256 & 112 & 352417\\
			\hline
			HMAC$\to$$K_{intermediate}$ & 256 & 112 & 352417\\
			\hline
			HMAC$\to$$K_{EAS-PSK}$ & 256 & 40 & 352417\\
			\hline
			HMAC$\to$$K_{E2E-enc}$ & 256 & 24 & 352417\\
			\hline
			HMAC$\to$$K_{E2E-int}$ & 256 & 24 & 352417\\			
			\hline
			\multicolumn{4}{|c|}{Data Transmission Phase}\\
			\hline
			AES-CBC/CMAC  & 128 & 128 & 3536+6219\\	
			\hline
	\end{tabular}}
\end{table}

\begin{table}[!htb]
	\renewcommand{\arraystretch}{1.3}
	\caption{computational cost for protocol based on SQN and AES}
	\label{SQN1_cost_overhead}
	\centering
	\resizebox{8.8cm}{!}{
		\begin{tabular}{|c|c|c|c|}
			\hline
			Operations & Key Size(bits) & Input Size(bits) & time(us) \\
			\hline
			\multicolumn{4}{|c|}{AKA Phase}\\
			\hline
			f1,f5(MAC,AK)  & 128 & 128 & 3515*2\\
			\hline
			HMAC$\to$$K_{AF}$ & 128 & 304 & 352417\\
			\hline
			HMAC$\to$$TID_{new}$ &128 & 256 & 352417\\
			\hline
			\multicolumn{4}{|c|}{Data Transmission Phase}\\
			\hline
			AES-CBC/CMAC  & 128 & 128 & 3536+6219\\	
			\hline
	\end{tabular}}
\end{table}

\begin{table}[!htb]
	\renewcommand{\arraystretch}{1.3}
	\caption{computational cost for protocol based on SQN and Ascon}
	\label{SQN2_cost_overhead}
	\centering
	\resizebox{8.8cm}{!}{
		\begin{tabular}{|c|c|c|c|}
			\hline
			Operations & Key Size(bits) & Input Size(bits) & time(us) \\
			\hline
			\multicolumn{4}{|c|}{AKA and Data Transmission Phase}\\
			\hline
			Ascon($C_{HN}$,$tag_{HN}$) & 128 & 304 & 12174\\
			\hline
			Ascon($C_{AIoT}$,$tag_{AIoT}$) & 128 & 128 & 12136\\
			\hline
	\end{tabular}}
\end{table}

\begin{table}[!htb]
	\renewcommand{\arraystretch}{1.3}
	\caption{computational cost for protocol based on nonce/PLK and AES}
	\label{nonce1_cost_overhead}
	\centering
	\resizebox{8.8cm}{!}{
		\begin{tabular}{|c|c|c|c|}
			\hline
			Operations & Key Size(bits) & Input Size(bits) & time(us) \\
			\hline
			\multicolumn{4}{|c|}{AKA Phase}\\
			\hline
			f1(MAC)  & 128 & 128 & 3515\\
			\hline
			HMAC$\to$$K_{AF}$ & 128 & 384 & 352417\\
			\hline
			HMAC$\to$$TID_{new}$ & 128 & 256 & 352417\\
			\hline
			\multicolumn{4}{|c|}{Data Transmission Phase}\\
			\hline
			AES-CBC/CMAC  & 128 & 128 & 3536+6219\\	
			\hline
	\end{tabular}}
\end{table}

\begin{table}[!htb]
	\renewcommand{\arraystretch}{1.3}
	\caption{computational cost for protocol based on nonce/PLK and Ascon}
	\label{nonce2_cost_overhead}
	\centering
	\resizebox{8.8cm}{!}{
		\begin{tabular}{|c|c|c|c|}
			\hline
			Operations & Key Size(bits) & Input Size(bits) & time(us) \\
			\hline
			\multicolumn{4}{|c|}{AKA and Data Transmission Phase}\\
			\hline
			Ascon($C_{HN}$,$tag_{HN}$) & 128 & 384 & 12193\\
			\hline
			Ascon($C_{AIoT}$,$tag_{AIoT}$) & 128 & 128 & 12136\\
			\hline
	\end{tabular}}
\end{table}

\begin{table}[!htb]
	\renewcommand{\arraystretch}{1.3}
	\caption{comparison of computational cost for all protocols}
	\label{cost_overhead}
	\centering
	\resizebox{8.8cm}{!}{
		\begin{tabular}{|c|c|}
			\hline
			Protocols & Computational Cost(s) \\
			\hline
			5G AKA  & 3.5 \\
			\hline
			EAP-AKA' & 5.3 \\
			\hline
			EPS AKA & 2.8\\
			\hline
			CP CIoT AKA & 2.1\\			
			\hline
			BEST AKA1 & 2.8\\			
			\hline
			BEST AKA2 & 3.2\\
			\hline
			BEST AKA3 & 2.5\\
			\hline
			Protocol (SQN \& AES) & 0.7\\
			\hline
			Protocol (SQN \& Ascon) & 0.02\\
			\hline
			Protocol (nonce/PLK \& AES) & 0.7\\
			\hline
			Protocol (nonce/PLK \& Ascon) & 0.02\\
			\hline
	\end{tabular}}
\end{table}

\begin{table}[!htb]
	\renewcommand{\arraystretch}{1.3}
	\caption{comparison of energy overhead for all protocols}
	\label{energy_cost_overhead}
	\centering
	\resizebox{8.8cm}{!}{
		\begin{tabular}{|c|c|c|c|c|c|c|}
			\hline
			\multirow{2}{*}{Protocols} & \multicolumn{3}{c|}{Power: 10$\mu$w} & \multicolumn{3}{c|}{Power: 300$\mu$w} \\
			 \cline{2-7}
		 	&   EO(10E-5J)    & EO/(5.4*10E-5J)    &  EO/(5.4*10E-3J)&  EO(10E-5J)    & EO/(5.4*10E-5J)    &  EO/(5.4*10E-3J)\\   
			 \hline
			5G AKA                     &    3.5 &  64.8\%   &   0.648\%   &    105   &   1944.4\%    &  19.4\% \\
			\hline
			EAP-AKA'                   & 5.3  &  98.1\%   &   0.981\%   &    159   &   2944.4\%    &  29.4\% \\
			\hline
			EPS AKA                    & 2.8  &  51.9\%   &   0.519\%   &    84   &   1555.6\%    &  15.6\% \\
			\hline
			CP CIoT AKA                & 2.1 &  38.9\%   &   0.389\%   &    63   &   1166.7\%    &  11.7\% \\		
			\hline
			BEST AKA1                  & 2.8&  51.9\%   &   0.519\%   &    84   &   1555.6\%    &  15.6\% \\		
			\hline
			BEST AKA2                  & 3.2&  59.3\%   &   0.593\%   &    96   &   1777.8\%    &  17.8\% \\
			\hline
			BEST AKA3                  & 2.5&  46.3\%   &   0.463\%   &    75   &   1388.9\%    &  13.9\% \\
			\hline
			Protocol (SQN \& AES) & 0.7&  13.0\%   &   0.130\%   &    21   &   388.9\%    &  3.9\% \\
			\hline
			Protocol (SQN \& Ascon) & 0.02&  0.37\%   &   0.0037\%   &    0.6   &   11.1\%    &  0.1\% \\
			\hline
			Protocol (nonce/PLK \& AES) & 0.7&  13.0\%   &   0.130\%   &    21   &   388.9\%    &  3.9\% \\
			\hline
			Protocol (nonce/PLK \& Ascon) & 0.02&  0.37\%   &   0.0037\%   &    0.6   &   11.1\%    &  0.1\% \\
			\hline
	\end{tabular}}
\end{table}

\subsection{Energy Overhead}

We estimate  the energy overhead (EO) of the above protocols  based on  the equation $W=P*t$, $W$ means that power consumption, $P$ means the operating electric power, $t$ means the executation time of the authentication protocol. Here we assume the AIoT devices are equipped with a  limited capacitor and  only evaluate the power consumption of the above protocols' security computations based on the previous experimental result.

Referring to  a fully charged capacitor of 36uAs@1.5V (i.e., $5.4*10^{-5}$J) and a solid-state battery of 1uAh@1.5V (i.e., $5.4*10^{-3}$J) used for AIoT devices in Annex B TR 22.840 \cite{22840}, considering that AIoT devices may have different operating power, we take 10$\mu$w and 300$\mu$w as examples, then we analyze the  power consumption of security calculation for one round terminal authentication and secure transmission implemented by AIoT devices based on different authentication protocols. 

 The comparison of energy overhead for all protocols  are listed in Table \ref{energy_cost_overhead}. 
For capacitor1, when the operating power is 10$\mu$w, the security power consumption of the existing protocols account for at least 38.9\% of the capacitor1. However, our proposed protocols have a maximum proportion of 13\% and a minimum of 0.37\%. The latter incurs the cost of requiring network systems to support new algorithms. Once the operating power increases to 300$\mu$w, the proportion of capacitor1's power consumption is at least 1166.7\%, making it difficult for this capacitor to support the complete execution of the protocol, except our proposed protocol based on Ascon, which is with a proportion of 11.1\%.

For capacitor2, when the operating power is 10$\mu$w, the security power consumption of the existing protocols accounts for less than 1\% of capacitor2. However, once the operating power increases to 300$\mu$w, the proportion of capacitor2's power consumption is at least 11.7\%. In contrast, our proposed protocol based on AES has a proportion of 3.9\%, and the protocol based on Ascon has a proportion of only 0.1\%. These results demonstrate our remarkable advantage.

\section{Conclusion}
In the future, it will be a inevitable trend for the AIoT applications, meanwhile attendant security issues will be non-negligible and critical in the wake of more application scenarios. However, due to the characteristic of AIoT devices, such as batter-free and ultra-low computing capability, the existing security protocols cannot be suitable for the kind of special device.
We propose  new  authenticated key agreement protocols based on traditional algorithm and novel Ascon algorithm. These protocols can provide strong security properties. In addition, we verify through simulation that the current protocols have a long running time and large power consumption, and our protocols are more energy efficient and suitable for AIoT devices.

% or


%\appendices

%\section{}

%\section*{Acknowledgment}

\ifCLASSOPTIONcaptionsoff
  \newpage
\fi

\end{document}